\documentclass[journal,twoside]{IEEEtran}

\usepackage{cite}
\usepackage{graphicx}
\usepackage{bm}

\def\mytitlestring{Finite-Difference and Pseudospectral Time-Domain
  Methods Applied to Backwards-Wave Metamaterials}
\begin{document}

\title{\mytitlestring}
\author{Michael~W.~Feise, 
  John~B.~Schneider,~\IEEEmembership{Member,~IEEE,}
  Peter~J.~Bevelacqua%
\thanks{Manuscript received April 17, 2003. Revised manuscript
  received \today.}
\thanks{This work was supported by Office of Naval Research code 321OA.}%
\thanks{Work performed at the School of Electrical Engineering and
  Computer Science, Washington State University, Pullman, WA 99164,
  USA (email: schneidj@eecs.wsu.edu).  MWF currently with Nonlinear
  Physics Group, Research School of Physical Sciences and Engineering,
  Australian National University, Canberra, ACT 0200, Australia.  PJB currently
  attends Stanford University.
}%
}
%
\markboth{ACCEPTED BY IEEE TRANSACTIONS ON ANTENNAS AND PROPAGATION
  FOR JANUARY 2005}{ACCEPTED BY IEEE TRANSACTIONS ON ANTENNAS AND
  PROPAGATION FOR JANUARY 2005}

\pubid{0018-926X/04\$XX.XX \copyright{} 2004 IEEE}

\maketitle

\begin{abstract}
  Backwards-wave (BW) materials that have simultaneously negative real
  parts of their electric permittivity and magnetic permeability can
  support waves where phase and power propagation occur in opposite
  directions.  These materials were predicted to have many unusual
  electromagnetic properties, among them amplification of the
  near-field of a point source, which could lead to the perfect
  reconstruction of the source field in an image [J.\ Pendry, Phys.\ 
  Rev.\ Lett.\ \textbf{85}, 3966 (2000)].  Often systems containing BW
  materials are simulated using the finite-difference time-domain
  technique.  We show that this technique suffers from a numerical
  artifact due to its staggered grid that makes its use in simulations
  involving BW materials problematic.  The pseudospectral time-domain
  technique, on the other hand, uses a collocated grid and is free of
  this artifact.
  It is also shown that when modeling the dispersive BW material, the
  linear frequency approximation method introduces error that affects
  the frequency of vanishing reflection, while the auxiliary
  differential equation, the Z transform, and the bilinear frequency
  approximation method produce vanishing reflection at the correct
  frequency.  The case of vanishing reflection is of particular
  interest for field reconstruction in imaging applications.
\end{abstract}


\IEEEpeerreviewmaketitle
\section{Introduction}
\label{sec:introduction}
\PARstart{E}{lectromagnetic} meta-materials with simultaneously
negative electric permittivity and magnetic permeability have recently
received much attention.  These materials can support a
backwards-traveling wave where the phase propagation is antiparallel
to the direction of energy flow.  These materials have been identified
by several names including left-handed or double-negative material but
we shall refer to them here as backwards-wave (BW) materials.  Their
properties were first considered theoretically by
Veselago\cite{Veselago:SPU-10-509} during the 1960's but they have only
been fabricated recently
\cite{Shelby:SCIENCE-292-1-77,Smith:PRL-84-4184}.  They are predicted
to exhibit many unusual properties such as refraction at a negative
angle, an inverse Doppler shift, and a backwards oriented \v{C}erenkov
radiation cone\cite{Veselago:SPU-10-509}.
Veselago\cite{Veselago:SPU-10-509} also predicted that, under the
conditions
\begin{equation}
  \label{eq:ideal-condition}
\varepsilon_r=\mu_r=-1,  
\end{equation}
a parallel slab of this material could focus waves emitted from a
point source and essentially act as a lens, albeit without
magnification.  These conditions are considered ``ideal'' because they
lead to the same impedance and speed of light as free space.

Pendry \cite{Pendry:PRL-85-3966} pointed out that under these
conditions incident decaying evanescent waves become growing inside
the slab.  The recovery of the evanescent waves allows the image
resolution to be better than the diffraction limit and this material
could theoretically produce a perfect image of the source, which lead
Pendry to call the system a ``perfect lens.''  A BW material must
necessarily be dispersive\cite{Veselago:SPU-10-509} and therefore all
this would only work properly at certain frequencies, where the
conditions of (\ref{eq:ideal-condition}) are fulfilled.  At these
frequencies the meta-material slab not only compensates for the phase
propagation of the waves, as a conventional lens does, but also
compensates for the amplitude decay experienced by evanescent waves.

This claim of growing evanescent waves and the possibility of a
perfect lens has aroused much interest and
given rise to much discussion of its correctness and feasibility, e.g.,
\cite{PRL:Hooft-87-249701,PRL:Pendry-87-249702,Garcia-PRL-88-207403,Gomez-Santos:PRL-90-077401,PRL:Valanju-88-187401,PRL:Pendry-90-029703,PRL:Valanju-90-029704,Grbic:APL-82-1815}.
Besides that, there is great interest and potential for applications
in these materials beyond the issue of a perfect lens.

In this paper we study the behavior of an evanescent wave interacting
with a slab of BW material via simulations.  We
consider the performance and applicability of two numerical techniques
being used to simulate this interaction.  In
Sec.~\ref{sec:model-system} we introduce the model system and in
Sec.~\ref{sec:numer-prop-schem} the numerical schemes.
Section~\ref{sec:results} discusses results and
Sec.~\ref{sec:conclusion} presents conclusions.
\pubidadjcol

\section{Model System}
\label{sec:model-system}
To be able to study the properties of evanescent waves interacting
with a BW slab, we isolate a single evanescent wave by inserting the
slab into a parallel plate waveguide and use excitation below the
cutoff frequency of the waveguide.  
This way the wave in the
waveguide is evanescent and its wave vector can be chosen through the
width of the waveguide and the spatial profile of the source.
\begin{figure}[tbp]
  \centerline{\includegraphics[width=3.5in]{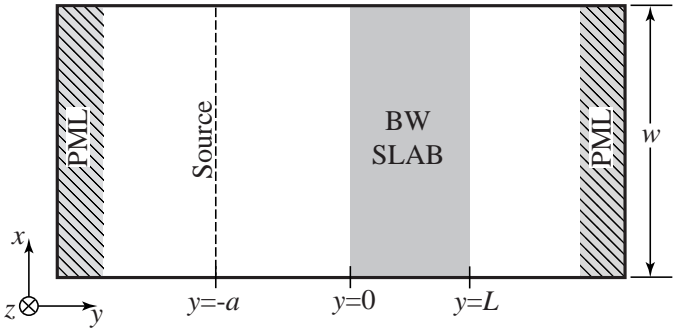}}
  \caption{Schematic view of the model system.  The parallel plate
    waveguide with width $w$ is translationally invariant in the $z$
    direction.  A sheet current source at $y=-a$ and a slab of BW
    material are located inside the waveguide. }
  \label{fig:schematic-waveguide}
\end{figure}
Figure~\ref{fig:schematic-waveguide} shows a schematic of the system.
The top and bottom of the waveguide are perfectly electrically
conducting (PEC) plates.  PEC walls also exist at the sides of the
computational domain but these are behind perfectly-matched layers
(PML) which essentially allow the structure to mimic an open waveguide
(but the evanescent fields considered here decay rapidly enough that
the performance of the PML is not a critical component in the
simulation).  The system is translationally invariant in the $z$
direction and can be treated as two-dimensional.  A sheet current
density $\mathbf{j}$ at $y=-a$ parallel to the $x$-$z$ plane acts as
the source for the fields.
As our interest is in the behavior of evanescent fields, a single
evanescent mode is excited using the spatial profile of a sheet
current matched to that of the lowest order evanescent mode,
\begin{equation}
  \label{eq:source-profile}
  \mathbf{j}(x,y,t)=\hat z j(t)
  \delta(y+a)\sin\left(\pi\frac{x}{w}\right),
\end{equation} 
where $\delta$ is the Dirac delta function and $w$ is the width of the
waveguide.  The source location and the waveguide width are shown in
Fig.~\ref{fig:schematic-waveguide}, and $j(t)$ is the time dependence
of the current source.
(In the steady-state situation we deal with a single or few wave
vectors.  For this reason the concern of
Ref.~\cite{Garcia-PRL-88-207403} does not apply here.)

Composite materials in which the effective electric permittivity and
the effective magnetic permeability are both negative in some
frequency interval have experimentally been shown to
exist\cite{Shelby:SCIENCE-292-1-77,Smith:PRL-84-4184}.  Based on this
we assume that both the electric permittivity and the magnetic
permeability are described by Lorentz-type frequency dependences as
\begin{eqnarray}
  \label{eq:Lorentz-material-e}
  \varepsilon_r(\omega) &=& 1 +
  \frac{\omega_{ep}^2}{\omega_{e0}^2 + i\gamma_e\omega -\omega^2},\\
  \label{eq:Lorentz-material-m}
  \mu_r(\omega) &=& 1 +
  \frac{\omega_{mp}^2}{\omega_{m0}^2 + i\gamma_m\omega -\omega^2}.
\end{eqnarray}
Here $\omega_{ep}$ and $\omega_{mp}$ are the plasma frequency,
$\omega_{e0}$ and $\omega_{m0}$ are the resonance frequency, and
$\gamma_e$ and $\gamma_m$ are the absorption parameter of the
permittivity and permeability, respectively.  To simplify the model we
choose $\varepsilon_r$ and $\mu_r$ to coincide, i.e.,
$\omega_{ep}=\omega_{mp}=\omega_p$,
$\omega_{e0}=\omega_{m0}=\omega_0$, and $\gamma_e=\gamma_m=\gamma$.
In the special case of $\omega_0=\gamma=0$, the conditions
(\ref{eq:ideal-condition}) are fulfilled at the design frequency
\begin{equation}
\label{eq:1}
f_d=\frac{\omega_p}{2\pi\sqrt{2}}.
\end{equation}
We take (\ref{eq:1}) to define the relationship between the design
frequency and the plasma frequency even when $\omega_0$ and $\gamma$
are nonzero but small.
                                
Our attention is primarily concerned with the design frequency since
this is the only frequency at which a perfect focus could possibly be
achieved.  At a frequency other than the design frequency the rate at
which evanescent fields grow in the slab is not equal to the rate they
decay in free space.  Thus for a source emitting multiple evanescent
fields in front of the slab, there is no unique point on the other
side of the slab at which the fields have obtained the same level as
at the source (i.e., there is no unique image point).

We assume that the material has abrupt edges and the change in
constitutive parameters is instantaneous and collocated.  The case of
the change in the constitutive parameters of a BW slab not coinciding
leads to a shift in the frequency of the nonreflecting wave and the
bound modes of the slab\cite{Feise:PRB-66-035113}.  
The fields of the bound modes decay exponentially away from the slab.
The two bound modes with spatially decaying fields inside and outside
the slab delimit a frequency interval that also contains a
nonreflecting wave with exponential spatial dependence.

For the material reaction to the field to be causal, the
Kramers-Kronig relations 
\cite{Kronig:JOSA-12-547,Kramers:ACIF-2-545} show that for any
deviation from free-space behavior the imaginary part of the
permittivity or permeability cannot vanish for all frequencies.  On
the other hand, as far as causality is concerned, it can be
arbitrarily small.

\section{Numerical Schemes}
\label{sec:numer-prop-schem}
The numerical schemes used here are applied directly to Maxwell's
equations rather than to a wave equation.  The first-order partial
differential equations are discretized in time and space and an update
algorithm is developed.  Specifically the behavior of the traditional
Yee finite-difference time-domain (FDTD) algorithm
\cite{Yee:TAP-14-302,Taflove:CED-1995} and the pseudospectral
time-domain (PSTD) method\cite{Liu:MOTL-15-158} are studied.  In both
schemes the system is modeled as two dimensional and the open sides of
the waveguide are modeled by using a causal uniaxial perfectly matched
layer (PML) absorbing boundary
\cite{Berenger:JCP-114-185,Sacks:TAP-43-1460,Muzuoglu:MGWL-6-447}.
One significant difference between the two schemes lies in the grid
location where the fields are sampled.  In the PSTD scheme the sample
points of the fields are collocated while in the FDTD scheme they are
staggered.

The plates of the parallel plate waveguide are modeled as perfect
electric conductors.  On the PEC boundary, the tangential
components of the electric field and the normal component of the
magnetic field must vanish.  The boundary conditions of the other
components are given by surface charge density $\rho_s$ and surface
current density $\mathbf{J}_S$,
\begin{eqnarray}
  \label{eq:2}
  \hat \mathbf{n}\times\mathbf{E} &=& \mathbf{0},\\
  \label{eq:3}
  \hat \mathbf{n}\cdot\mathbf{E} &=& \frac{\rho_s}{\varepsilon},\\
  \label{eq:4}
  \hat \mathbf{n}\times\mathbf{H} &=& \mathbf{J}_S,\\
  \label{eq:5}
  \hat \mathbf{n}\cdot\mathbf{H} &=& 0.
\end{eqnarray}
Here $\hat \mathbf{n}$ is the surface normal unit vector.

\subsection{Yee Scheme}
\label{sec:Yee-scheme}
In 1966, Yee\cite{Yee:TAP-14-302} proposed a discretization scheme for
Maxwell's equations based on a fully staggered grid, where the electric
field components are each sampled on one of the edges of a cubical
unit cell and the magnetic components each on one of the face centers.
In this manner the spatial derivatives can be approximated as centered
differences and one achieves second-order accuracy in the spatial step
size\cite{Taflove:CED-1995}.

The PEC boundary conditions are implemented by placing the boundary in
a plane that contains the tangential electric field sampling points
and then setting these components to zero.  In this case, the
tangential magnetic field components and the normal electric field
component are not sampled on the boundary and one does not need to
enforce any boundary condition on them.  The normal component of the
magnetic field is sampled on the boundary and is updated in the usual
fashion.  Its boundary condition is automatically fulfilled through
its dependence on the tangential electric field.

Owing to the staggering of the field components, the precise location
of an interface where both the permittivity and permeability change is
ambiguous.  The case where a single material parameter, such as the
permittivity, changes has been well-studied in the literature, e.g.,
\cite{Celuch:EMC-1994,Hwang:MWCL-11-2001}.  It is well established
that for a tangential field component collocated with the interface,
the best approach is to use the arithmetic average of the material
properties to either side.  (It can also be shown, at least for the
case of normal incidence, that if one is free to assign the material
interface to fall halfway between tangential nodes, an abrupt change
in the material parameters is optimum and, in fact, superior to
averaging \cite{Schneider:URSI-2000}.)  For the study here, the
precise location of the interface is not a primary concern.  Instead,
the important issue is whether the evanescent fields exhibit the
proper growth within the BW slab (whatever its thickness or location).
As reported in \cite{Ziolkowski:PRE-64-056625}, averaging of material
parameters has been used for modeling BW slabs with the Yee
algorithm.  (In that work surface waves are present and no
subwavelength imaging is seen.)

\begin{figure}[tbp]
  \centerline{\includegraphics[width=3.5in]{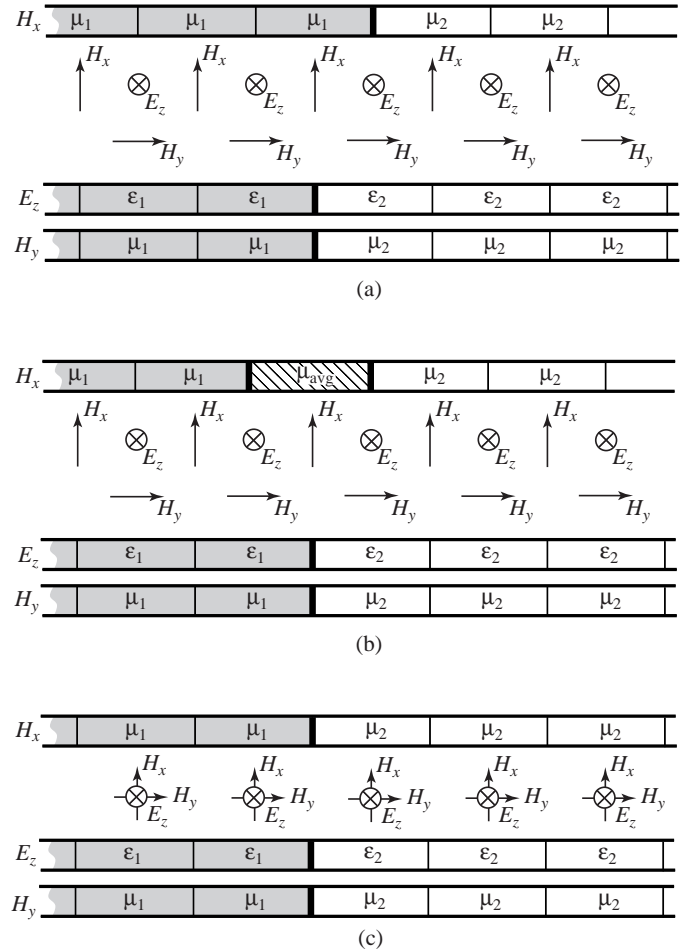}}
  \caption{Depiction of the way the permittivity and permeability
  change at an interface.  The horizontal bars show the material
  parameters associated with a particular field component (the
  associated field
  is indicated by the label to the left of the bar).  A slice of the
  grid indicating the node positions is given for each of the cases.
  (a) The Yee grid with ``abrupt'' boundaries.  The interface is
  assumed to exist 
  between the last $H_x$ which uses
  permeability $\mu_1$ and the first $E_z$ node which uses
  permittivity $\epsilon_2$. (b) The Yee grid using averaging.  The
  interface is assumed to be collocated with the $H_x$ node which uses
  the average permittivity to either side even though the introduction
  of a new material ($\mu_{\mathrm{avg}}$) arguably creates an additional
  interface further to the left.  (c) The PSTD grid.  Because the grid
  is not staggered, the boundary between the two regions is
  unambiguously assumed to exist between the nodes which employ the
  different material parameters.}  \label{fig:boundaries}
\end{figure}

Figure \ref{fig:boundaries} shows a portion of the computational grid
as well as the material properties associated with each node---one must
attribute a permeability to each magnetic field node and a
permittivity to each electric field node.  Figure
\ref{fig:boundaries}(a) shows the case of using abrupt discontinuities
in the material parameters in the Yee grid while (b) shows the case
where averaging is used for the tangential node at the interface.
Figure \ref{fig:boundaries}(c) shows the grid for the PSTD simulation
which is described next.

\subsection{Pseudospectral Scheme}
\label{sec:PSTD-scheme}
In pseudospectral techniques a discrete Fourier transform is used to
calculate the spatial derivatives in the discretized version of a
partial differential equation.  Time integration is achieved using the
same second-order time-stepping approach as used in the Yee algorithm.
Unlike the Yee algorithm where central-differences are used, in PSTD
the spatial derivatives are calculated at the same location as the
field.  As shown in Fig.\ \ref{fig:boundaries}(c), for Maxwell's
equations this means that the electric and the magnetic field
components are all sampled at the same location, i.e., in the center
of a cubic unit cell. An important problem with this approach is that
the use of exponential Fourier transforms inherently introduces
periodic boundary conditions that are often undesired.
This problem can be avoided by covering the numerical domain
boundaries with a PML layer such that the waves are absorbed in the
PML before they can contaminate the simulation through the periodic
boundaries\cite{Liu:MOTL-15-158}.  This gives one a tool to simulate
open domains.  If the system contains PEC boundaries, the problem
becomes much more difficult.  Nonetheless, in a two-dimensional system
in transverse electric (TE) polarization, i.e., $\mathbf{E}$
perpendicular to the plane, which is source-free near the boundaries,
one can show that the PEC boundary conditions
(Eqs.~\ref{eq:2}--\ref{eq:5}) imply that
\begin{equation}
  \label{eq:6}
\left(\hat \mathbf{n}\cdot\mathbf{\nabla}\right)\left(\hat \mathbf{n} \times\mathbf{H} \right) =\mathbf{0}.
\end{equation}
Here $\hat \mathbf{n}$ is the surface normal unit vector and $\hat
\mathbf{n}\cdot\mathbf{\nabla}$ is the derivative in that direction.
Conditions (\ref{eq:2}) and (\ref{eq:6}) can be easily enforced with
Fourier sine and cosine transforms if one chooses Fourier sine
transforms for the tangential electric field components and Fourier
cosine transforms for the tangential magnetic field components.  No
spatial derivatives of normal components appear and no boundary
condition needs to be enforced on them.  Thus, in a two-dimensional
system in TE polarization, the PEC boundary conditions are
conveniently implemented through the choice of Fourier transform.

The discrete Fourier transform has problems representing a delta
function, discretized as a Kronecker delta, correctly.  For this
reason it is advantageous to use a spatially smoothed rather than a
highly localized source \cite{Liu:MOTL-15-158} to represent the source
screen of (\ref{eq:source-profile}).  This problem also appears when a
sudden change in material parameters leads to a sudden change in the
fields.  In the simulations one observes wiggles or ripples on the
signal that can become significant if the change in field is strong.
This artifact is the Gibbs phenomenon associated with the Fourier
transforms.

\subsection{Dispersion Implementation}
\label{sec:dispersion-implementation}
To achieve negative $\varepsilon$ and negative $\mu$ in a
material, this material must necessarily be dispersive
\cite{Veselago:SPU-10-509}.  In our model system we chose Lorentz
dispersion characteristics
(\ref{eq:Lorentz-material-e},\ref{eq:Lorentz-material-m})
\cite{Shelby:SCIENCE-292-1-77,Pendry:PRL-85-3966}.  There are several
commonly used methods to implement frequency dependence in a
time-domain algorithm, e.g., the auxiliary differential equation (ADE)
method\cite{Joseph:OL-16-1412}, the recursive convolution method
\cite{Luebbers:TAP-39-29,Kelley:TAP-44-792}, and the Z transform
method
\cite{Sullivan:TAP-40-1223,Sullivan:TAP-44-28,Sullivan-EMS-2000}.  In
this study we use the ADE method, the Z transform method, and two
frequency approximation methods\cite{Hulse:JOSAA-11-1802}.  Here we
present the update for the electric field; analogous equations hold
for the magnetic field.  In the following, $\Delta_t$ will denote the
temporal step size of the discretization scheme.

\subsubsection{Auxiliary Differential Equation Method}
\label{sec:auxill-diff-equat}
One approach to implement a frequency dependent response in a discrete
time-domain method is to transform the frequency-domain constitutive
parameters into the time domain and then approximate the derivatives
with differences based on Taylor series expansions
\cite{Joseph:OL-16-1412}.
The frequency dispersion (\ref{eq:Lorentz-material-e}) at any given
electric field node inside the material slab is implemented as
\begin{eqnarray}
  \label{eq:7}
  E^n \hspace{-0.095in} &=& \hspace{-0.095in} 
         \left\{ \left(\omega_0^2\Delta_t^2+\gamma\Delta_t+2\right)
         \frac{D^n}{\varepsilon_0} -4  \frac{D^{n-1}}{\varepsilon_0}
         \right. \nonumber\\
  \hspace{-0.095in} && \hspace{-0.095in} \mbox{}
    + \left(\omega_0^2\Delta_t^2-\gamma\Delta_t+2\right)
    \frac{D^{n-2}}{\varepsilon_0}\nonumber\\
  \hspace{-0.095in} && \hspace{-0.095in} \mbox{}
    + \left. \rule[-5pt]{0pt}{20pt}
    4 E^{n-1} -\left[\left(\omega_p^2+\omega_0^2\right)
    \Delta_t^2-\gamma\Delta_t+2\right] E^{n-2} \right\}\nonumber\\
  \hspace{-0.095in} && \hspace{-0.095in} \mbox{}
    \times \left[\left(\omega_p^2+\omega_0^2\right)
    \Delta_t^2+\gamma\Delta_t+2\right]^{-1},
\end{eqnarray}
where $D$ is the electric flux density which must be stored as a
separate quantity.  The superscripts on the fields denote the time
step at which the field values are taken.

\subsubsection{Z Transform Method}
\label{sec:z-transform-method}
The Z transform is essentially a more general version of the
discrete Fourier transform.  It is strictly based on a sampled signal
and provides an exact transformation between sample domain and Z
domain.  For frequency dependencies of the Debye, plasma, and Lorentz
type and even some types of non-linearities it allows one to derive
exact update algorithms (i.e., exact in the sampled sense) for the sampled signal
\cite{Sullivan:TAP-40-1223,Sullivan:TAP-44-28,Sullivan-EMS-2000,Hulse:JOSAA-11-1802}.
The frequency dispersion of (\ref{eq:Lorentz-material-e}) is
implemented as
\begin{eqnarray}
  \label{eq:8}
  E^n \hspace{-0.095in} &=& \hspace{-0.095in}
    \frac{D^n}{\varepsilon_0} - \frac{\omega_p^2}{\beta}\Delta_t I^{n-1},\\
  I^n \hspace{-0.095in} &=&\hspace{-0.095in} 
    2 e^{-\alpha\Delta_t}\cos\left(\beta\Delta_t\right)
    I^{n-1} - e^{-2\alpha\Delta_t} I^{n-2}\nonumber\\
  \hspace{-0.095in} &&\hspace{-0.095in} \mbox{}
    + e^{-\alpha\Delta_t}\sin\left(\beta\Delta_t\right) E^n,
\end{eqnarray}
where $\alpha=\gamma/2$, $\beta=\sqrt{\omega_0^2-\gamma^2/4}$, and $I$
is an auxiliary field.  This implementation is applicable for real
$\beta$, i.e., when the system response function is a damped sine wave.  When
$\beta$ is imaginary, the character of the response function changes
and the Z transform has to be implemented differently
\cite{Sullivan-EMS-2000}.

\subsubsection{Frequency Approximation Method}
\label{sec:freq-approx-method}
Another common method to derive discrete time-domain update equations
for the frequency dependence of a system is frequency approximation.
Here one starts with the frequency domain expression for the system
response and makes the transition to the discretized time domain by
replacing each occurrence of $i\omega$ by $\frac{1-z^{-1}}{\Delta_t}$,
i.e., a linear backward difference where $z^{-1}$ represents a shift
back in time of one temporal step
\cite{Hulse:JOSAA-11-1802,Sullivan-EMS-2000}.  When the frequency
dependence becomes complicated this method can make the derivation of
update equations much easier than the Z transform method.  On the
other hand, in general, while it does not significantly change the
computational burden, it introduces new error.  A bilinear
approximation, $i\omega\rightarrow\frac{2}{\Delta_t}
\frac{1-z^{-1}}{1+z^{-1}}$, gives better accuracy but increases the
computational burden \cite{Sullivan-EMS-2000,Hulse:JOSAA-11-1802}.
The frequency dispersion (\ref{eq:Lorentz-material-e}) using the
linear approximation is implemented as
\begin{eqnarray}
  \label{eq:9}
  E^n \hspace{-0.095in} &=& \hspace{-0.095in} 
    \left\{\left(\omega_0^2\Delta_t^2+\gamma\Delta_t+1\right)
    \frac{D^n}{\varepsilon_0}\right. \nonumber\\
  \hspace{-0.095in} && \hspace{-0.095in} \mbox{}
    - \left.  \rule[-5pt]{0pt}{20pt}
      \left(\gamma\Delta_t+2\right) I^{n-1} + I^{n-2}\right\}\nonumber\\
  \hspace{-0.095in} && \hspace{-0.095in} \mbox{}
    \times \left[\left(\omega_p^2+\omega_0^2\right)\Delta_t^2
    + \gamma\Delta_t +1\right]^{-1},\\ \label{eq:10}
  I^n \hspace{-0.095in} &=& \hspace{-0.095in} 
    \left\{\omega_p^2\Delta_t^2E^n +\left(\gamma\Delta_t+2\right) I^{n-1}
    -I^{n-2}\right\} \nonumber\\
  \hspace{-0.095in} && \hspace{-0.095in} \mbox{}
    \times \left[\omega_0^2\Delta_t^2 +\gamma\Delta_t +1\right]^{-1}.
\end{eqnarray}
In the bilinear approximation the electric field is updated using
\begin{eqnarray}
  \label{eq:11}
  E^n \hspace{-0.095in} &=& \hspace{-0.095in}
    \left\{ \left(\omega_0^2\Delta_t^2+2\gamma\Delta_t+4\right)
    \frac{D^n}{\varepsilon_0} -2\omega_p^2\Delta_t^2 E^{n-1} \right. \nonumber\\
  \hspace{-0.095in} && \hspace{-0.095in} \mbox{} 
      + \left. \rule[-5pt]{0pt}{20pt}
      \left(2\omega_0^2\Delta_t^2-8\right) I^{n-1} +
      \left(\omega_0^2\Delta_t^2-2\gamma\Delta_t+4\right) I^{n-2} \right\}
      \nonumber\\
  \hspace{-0.095in} && \hspace{-0.095in} \mbox{} 
     \times \left[ \left(\omega_p^2+\omega_0^2\right)\Delta_t^2
     +2\gamma\Delta_t+4\right]^{-1},\\ \label{eq:12}
  I^n \hspace{-0.095in} &=& \hspace{-0.095in} 
     \left\{\omega_p^2\Delta_t^2 E^n +2\omega_p^2\Delta_t^2 E^{n-1}
     + \omega_p^2\Delta_t^2 E^{n-2}\right. \nonumber\\
  \hspace{-0.095in} && \hspace{-0.095in} \mbox{}
     -\left. \left(2\omega_0^2\Delta_t^2-8\right) I^{n-1} 
     -\left(\omega_0^2\Delta_t^2-2\gamma\Delta_t+4\right)I^{n-2} \right\}
     \nonumber\\
  \hspace{-0.095in} && \hspace{-0.095in} \mbox{}
     \times \left[ \omega_0^2\Delta_t^2+2\gamma\Delta_t+4 \right]^{-1}.
\end{eqnarray}
In (\ref{eq:8})--(\ref{eq:12}) $I$ is used as an auxiliary field.

\section{Results}
\label{sec:results}
We have performed model calculations for a slab of BW material inside
a waveguide using the geometry and the material described in
Sec.~\ref{sec:model-system}.  We compare the results for the FDTD and
the PSTD schemes, using the ADE method for the dispersion
implementation.  We also show a comparison of all four dispersion
implementations of Sec.~\ref{sec:dispersion-implementation} with the
PSTD scheme.

For the FDTD method one commonly uses the Yee
scheme\cite{Yee:TAP-14-302} to derive the update equations for the
fields.  Because this scheme is based on a staggered grid, where all
field components are sampled at different locations in a unit cell, it
has inherent difficulties representing a collocated change in material
parameters.
This problem becomes clearly evident when modeling an interface
between free space and a BW material.

In the PSTD method all the field components are sampled at the same
location and no transition layer exists.  We find that, unless the
dispersion implementation introduces phase error, the bound modes and
the nonreflecting wave appear at frequencies very close to those found
in the continuous world.

In the simulations the design frequency $f_d$ is $15\,\mbox{GHz}$ and
the spatial discretization level is 100 points per propagating
free-space wavelength at the design frequency $\lambda_f$, i.e.,
$\Delta_x=\Delta_y=\lambda_f/100=c/(100 f_d)=0.1999\,\mbox{mm}$.
Using such a fine discretization is typically unnecessary when
simulating propagating waves, where it is more customary to use
approximately 20 points per wavelength, but because evanescent waves
have rapid amplitude decay and also phase variations (transverse to
the amplitude decay) that have wavelengths smaller than the
propagating free-space wavelength, it can be necessary to use a finer
discretization to model these waves accurately.

The slab length $L$ is $33\Delta_y$, and the source is located a
distance $a=30\Delta_y$ in front of the BW slab.  The waveguide width
$w$ which determines the degree of evanescence was chosen to be
$32\Delta_x$.  (For $w>\lambda_f/2=50\Delta_x$ the waves in the
waveguide would be propagating.)  The material parameters were chosen
as $\omega_p/(2\pi)=\sqrt{2}f_d$, $\omega_0/(2\pi)=5\,\mbox{MHz}$, and
$\gamma/(2\pi)=5\,\mbox{MHz}$.  The values of $\omega_0$ and $\gamma$
are small compared to $\omega_p$ and their influence on the results is
weak.  Even though (\ref{eq:ideal-condition}) can only be
approximately fulfilled, due to the loss and the nonzero resonance
frequency, we will assume that the design frequency (\ref{eq:1}) is
unchanged.  The results for the Lorentz material are similar to those
obtained for an unmagnetized plasma, i.e., $\omega_0=0\,\mbox{Hz}$.
The time dependence of the source current is a sinusoid with an
exponential ramp $j(t)=(1-e^{-t/\tau})\sin(2\pi f_s t)$, with carrier
frequency $f_s$ and the ramping governed by $\tau$.  In the
simulations we used $\tau=5000\Delta_t$.

The graphs presented in this section show the spatial dependence of
the magnitude of a certain frequency component of the temporal Fourier
transform of the fields in the $y$ direction.  The temporal Fourier
transform was recorded between 15000 time steps and when the
simulation stopped at a total of 67000 time steps.  The solid vertical
lines indicate the extent of the BW slab, the dotted vertical line
indicates the source location, and the arrows show an arbitrarily
chosen object location and its corresponding image location.  Ideally,
at the design frequency the BW slab compensates exactly for the
free-space propagation and decay of incident waves corresponding to
the thickness of the slab.  Thus, referring to the geometry of
Fig.~\ref{fig:schematic-waveguide}, an object located at $y=-b$ with
$b\leq L$, is imaged at $y=2L-b$.

\subsection{Nonreflecting Wave}
\label{sec:results-nonreflecting-wave}
In the Yee discretization scheme, all field components are sampled at
different locations in a unit cell.  The permittivity and permeability
are usually also sampled at these locations.  A change in material
parameters at the interface between two materials, therefore, is
extended in space and the algorithm essentially sees a thin transition
layer, where some parameters are those of one material and the others
are those of the other material.  As discussed in
Sec.~\ref{sec:Yee-scheme}, the interface may be approximated with or
without an average of the material properties to either side.  We
identify the approach without averaging as the ``abrupt'' boundary.

For a BW slab, the effect of such a transition layer between the
change in $\varepsilon$ and the change in $\mu$ has previously been
studied using a continuous-space model \cite{Feise:PRB-66-035113}.
Our Yee simulations employing abrupt boundaries have transition layers
with the $\varepsilon$ of the BW slab and the $\mu$ of the surrounding
free space, similar to the situation in \cite{Feise:PRB-66-035113}.
For the front face of the slab, this corresponds to the situation
shown in Fig.\ \ref{fig:boundaries}(a) if one associates the subscript
1 with free space and the subscript 2 with the BW material.  When
averaging is used, the average permeability is applied to the $H_x$
nodes assumed to coincide with the interfaces.

\begin{figure}[tbp]
  \centerline{\includegraphics[width=3.5in]{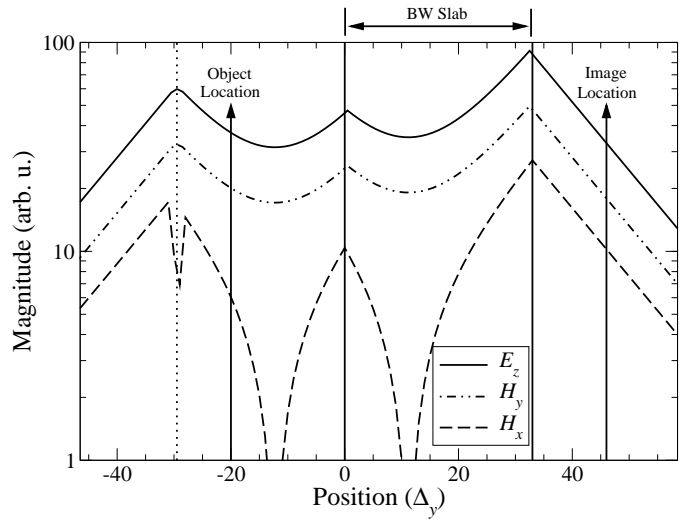}}
  \caption{Magnitude of the Fourier transform component at the design
  frequency using the abrupt-boundary Yee scheme with the ADE
  dispersion implementation.  The lines for the different fields have
  been normalized and offset to allow qualitative comparison.  The
  system parameters are given in the text.  The solid vertical lines
  indicate the extent of the BW slab, the dotted vertical line
  indicates the source location and the arrows show an arbitrarily
  chosen object location and its corresponding image location.}
  \label{fig:Yee-design-freq}
\end{figure}
\begin{figure}[tbp]
  \centerline{\includegraphics[width=3.5in]{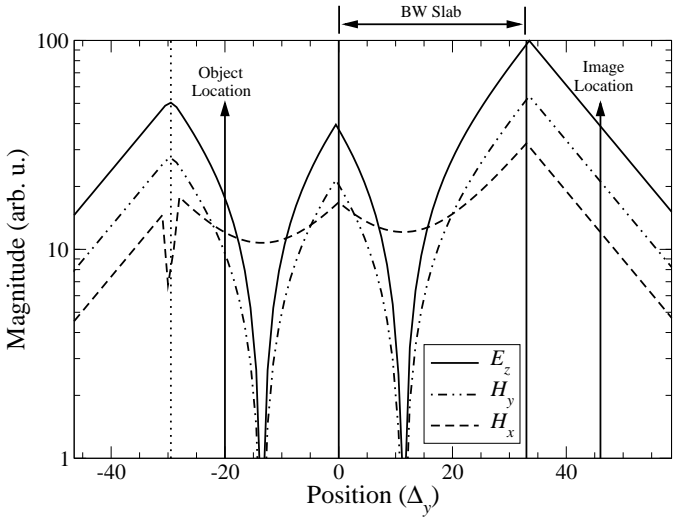}}
  \caption{Same as the previous figure except except averaging is used
  for the permeability at the interface.}
  \label{fig:Yee-design-avg}
\end{figure}
To illustrate the effect of the transition layer in the Yee scheme, we
show in Fig.~\ref{fig:Yee-design-freq} the fields at the design
frequency, where the source frequency $f_s$ equals $f_d$, using abrupt
boundaries and the ADE dispersion implementation.  The evanescent
field grows inside the BW slab but a reflected field is clearly
present, as can be seen from the change in slope of the fields between
the source and the $y=0$ surface, as well as between the two slab
surfaces.  In Fig.~\ref{fig:Yee-design-freq} the field amplitudes at
the image location are significantly different from those at the
object location.  Figure~\ref{fig:Yee-design-avg} again shows the
results at the design frequency except now averaging of the
permeability is used at the interface.  As before, the results do not
exhibit the desired behavior.  The deep nulls in both these figures
occur when the total field changes sign because the reflected field is
of equal magnitude as the field of the source at that point but has opposite
sign.

\begin{figure}[tbp]
  \centerline{\includegraphics[width=3.5in]{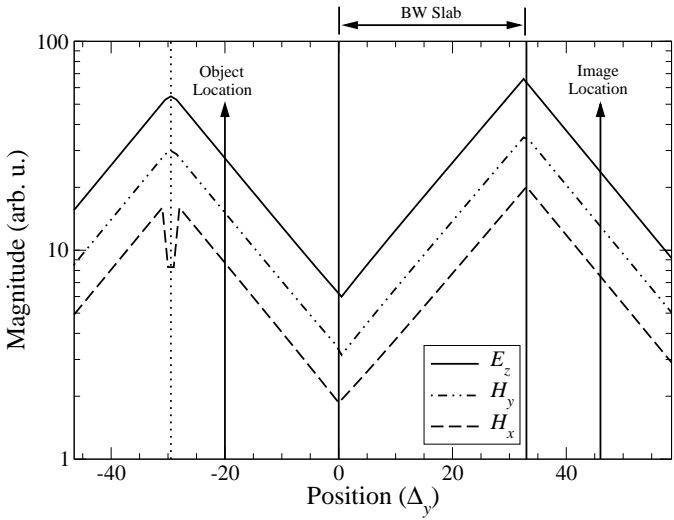}}
  \caption{Magnitude of the Fourier transform component at
  $f=14.83\,\mbox{GHz}$, away from the design frequency
  $f_d=15\,\mbox{GHz}$, using the abrupt-boundary Yee scheme with the
  ADE dispersion implementation.  The lines for the different fields
  have been normalized and offset to allow qualitative comparison.
  The system parameters are given in the text.}
  \label{fig:Yee-nonreflecting}
\end{figure}
Figure~\ref{fig:Yee-nonreflecting} shows the nonreflecting wave in the
Yee scheme with the abrupt boundary.  It occurs at 
$14.83\,\mbox{GHz}$, i.e., below the design frequency $f_d$ of
$15\,\mbox{GHz}$, and was found by scanning through the frequency
range between the two bound mode frequencies.  The field is clearly
amplified exponentially inside the BW slab and decays exponentially
outside of it.  Upon close inspection one finds that the magnitude of
the electric field at the object location is approximately 28, while
it is about 26 at the image location.  This discrepancy is due to the
slower speed of light in the BW material at this frequency.  Thus, the
exponential growth inside the slab is slower than the decay outside of
it and, though the field is amplified in the slab, the amplification
does not accurately compensate for the decay in free space.  The
magnitude of this discrepancy as well as the frequency at which the
nonreflecting wave occurs is dependent on the wavevector of the field
\cite{Feise:PRB-66-035113}.
Thus, it is not possible to define a single corrected image location
that holds for all wavevectors (this is true of both the propagating
and evanescent components).

\begin{figure}[tbp]
  \centerline{\includegraphics[width=3.5in]{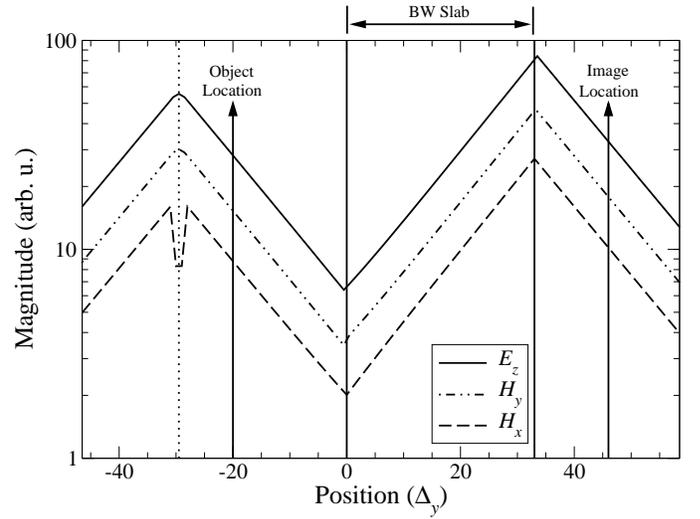}}
  \caption{Magnitude of the Fourier transform component at
  $f=15.17\,\mbox{GHz}$ using the Yee scheme with material averaging
  and the ADE dispersion implementation. The lines for the different fields
  have been normalized and offset to allow qualitative comparison.
  The system parameters are given in the text.}
  \label{fig:yeeAvgNRW}
\end{figure}

Figure~\ref{fig:yeeAvgNRW} shows the nonreflecting wave when averaging
is used for the boundary.  Here the nonreflecting wave was found to
occur at $15.17\,\mbox{GHz}$, i.e., above the design frequency.  An
object of field strength 28 was found to have a corresponding field
strength at the image point of 33 (i.e., the field was overcompensated).

In \cite{Feise:PRB-66-035113} it was found that transition layers
shift the frequency of vanishing reflection coefficient upward from
the design frequency.  In the FDTD simulations we find that for the
abrupt-boundary Yee scheme the nonreflecting frequency is shifted
downward from the design frequency.  We attribute this
difference to the effect the discretization has on the transmission
and reflection coefficient of each interface.  (This is in addition to
introducing transition layers into the system.)  When averaging is
used, the nonreflecting frequency is above the design frequency, but
the offset between the two frequencies is approximately the same as
when the abrupt boundary is used.  Thus there does not appear to be an
obvious advantage to using averaging when modeling a BW slab since
neither yields the correct behavior at the design frequency.
(However, again note that we are not concerned with the specific
location of the boundary, but rather just the behavior of the
evanescent fields.)

\begin{figure}[tbp]
  \centerline{\includegraphics[width=3.5in]{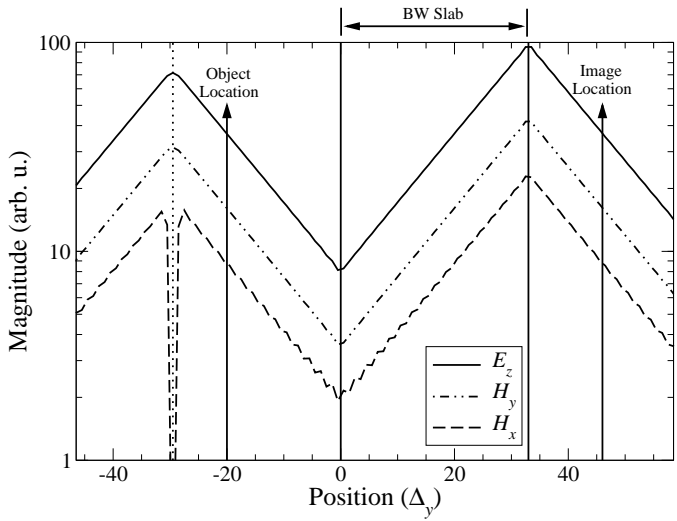}}
  \caption{Magnitude of the Fourier transform component at the
    design frequency using the PSTD scheme with the ADE dispersion
    implementation.  The lines for the different fields have been
    normalized and offset to allow qualitative comparison.  The system
    parameters are given in the text.  The ripple on $H_x$ is a
    numerical artifact of the Fourier transform due to finite spatial
    resolution.}
  \label{fig:PSTD-nonreflecting}
\end{figure}
In the PSTD scheme all field components are collocated and there is no
transition layer at a material interface.  Using the ADE method to
implement the frequency dispersion of the BW material we observe that
the nonreflecting wave occurs at the design frequency, as shown in
Fig.~\ref{fig:PSTD-nonreflecting}.  Because the wave vector magnitudes
inside and outside the material are nearly equal (with slight
discrepancy due to loss and a nonzero resonance frequency), the
amplification of the evanescent wave inside the slab compensates for
its free-space decay.  The field amplitudes at the object location and
its image location are within 0.5\% of each other.

The examples in Figs.~\ref{fig:Yee-design-freq}--\ref{fig:yeeAvgNRW}
are at a discretization level of $\lambda_f/100$.  Arguably, for
either the averaged or abrupt boundary, this provides a transition
layer thickness of only $\lambda_f/200$ in the Yee scheme and yet it
still produces a significant enough frequency shift for the
nonreflecting wave that realization of a perfect planar lens is
essentially impossible.  The analysis in \cite{Feise:PRB-66-035113}
shows that in the limit of vanishing layer thickness the frequency of
the nonreflecting wave goes to the design frequency in the continuous
case.  Nevertheless, for all practical discretization levels the
influence of the transition layer remains significant.  There have
been reports of the Yee scheme being used to model evanescent fields
in BW materials \cite{Rao:PRB-68-113103,Rao:PRE-68-067601}, but the
level of discretization that had to be used was on the order of 566 cells per
propagating wavelength \cite{Rao:personal} which may be prohibitive
for most applications.

In general all finite difference schemes suffer numerical dispersion.
The PSTD scheme is more computationally expensive than the Yee
algorithm for the same level of discretization but suffers less
numerical dispersion.  The dispersive properties of the PSTD scheme
have been discussed in the literature \cite{Liu:MOTL-15-158} (although
the dispersion relation given in the literature only pertains to the
case of an infinite grid and thus only holds approximately for finite
grids).  The dispersive properties for evanescent fields in FDTD
simulations have also be discussed in the literature
\cite{Schneider:MTT-49-280}.  The fields being considered here are
discretized such that numeric dispersion is not a significant concern.
Thus the major difference between the schemes considered here is the
treatment of the interface and one cannot attribute the superior
results of the PSTD scheme to its superior dispersion.

\subsection{Comparison of Dispersion Implementations}
\label{sec:comp-disp-impl}
\begin{figure}[tbp]
  \centerline{\includegraphics[width=3.5in]{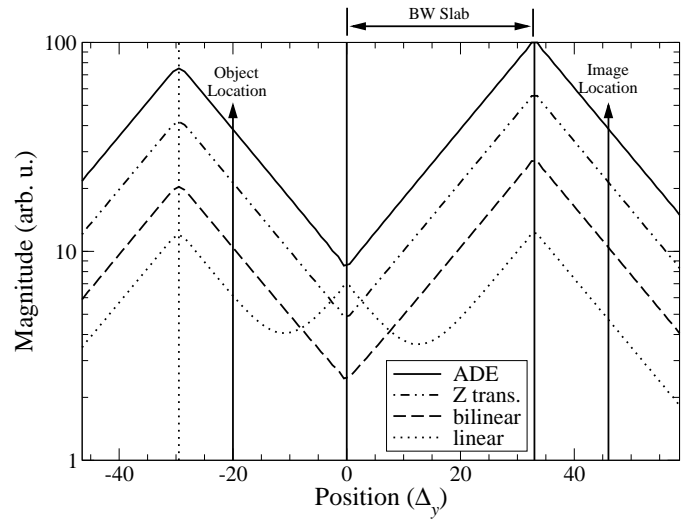}}
\caption{Magnitude of the Fourier transform component at the design
  frequency using the PSTD scheme.  The field is calculated using the
  ADE, Z transform, and the bilinear and the linear frequency
  approximation methods.  A reflected wave is clearly visible with the
  linear frequency approximation. The system parameters are given in
  the text.  The lines for the different fields have been offset to
  allow qualitative comparison.}
\label{fig:compare-dispersion-PSTD}
\end{figure}

In Sec.~\ref{sec:dispersion-implementation} we discussed four
different implementations of the frequency dependence of the
permeability and permittivity.  To compare their performance, we use
each one in a simulation using the PSTD scheme at the design
frequency, with the parameters given at the beginning of
Sec.~\ref{sec:results}.  In this configuration one expects the
reflected wave at the first interface ($y=0$) to vanish, as shown in
Fig.~\ref{fig:PSTD-nonreflecting}.  In
Fig.~\ref{fig:compare-dispersion-PSTD} we show the electric field in
these simulations.  It is apparent that the ADE, the Z transform, and the 
bilinear frequency approximation all give good results.  No reflected
wave is visible and upon inspection one finds that the magnitudes at
the object and image location are essentially equal.  When plotted
without offset the three curves fall on top of one another.  The linear
frequency approximation on the other hand does show a significant
amplitude for the reflected wave.  The field at the image location is
also much weaker than at the object location.  This effect is probably
due to the numerical phase error inherent in this method
\cite{Hulse:JOSAA-11-1802}.  This shows that the use of the linear
frequency approximation is problematic in simulations involving BW
materials, while many of the other common dispersive material implementations
work well.

\section{Conclusion}
\label{sec:conclusion}
We have studied the finite-difference time-domain and the
pseudospectral time-domain methods in the context of evanescent waves
in the presence of a backward-wave material.  Our simulations verify
that the predicted growth of the evanescent field inside the BW
material, which is vital for imaging beyond the diffraction limit,
does occur.  We found that the Yee FDTD method suffers from a
transition layer effect due to the inherent staggered grid, even when
using an averaging technique at the material interface, while the
collocated grid used in the PSTD method avoids this numerical
artifact.  The choice of numerical technique to implement the
frequency dependence of the permittivity and the permeability is also
important for the correct modeling of the BW material.

Thus, we have shown that the PSTD method in concert with the ADE
technique, the Z transform technique, or the bilinear frequency
approximation, allows one to accurately model the interaction of a BW
slab with an evanescent wave.

\section*{Acknowledgment}
The authors would like to thank Christopher L.\ Wagner of Washington
State University for useful discussion.

\bibliographystyle{IEEEtran}

\begin{biographynophoto}{Michael W.\ Feise}
  received the M.S.\ and the Ph.D.\ degree in Physics from Washington
  State University. From 2001 to 2003 he was a research associate with
  the School of Electrical Engineering and Computer Science at
  Washington State University.  In 2003 he became a research fellow in
  the Nonlinear Physics Group of the Research School of Physical
  Sciences and Engineering at the Australian National University. His
  research interests include electromagnetic and acoustic wave
  propagation and their linear and nonlinear interaction with matter,
  as well as THz radiation and low-dimensional semiconductor devices.
\end{biographynophoto}

\begin{biographynophoto}{John B.\ Schneider}
  received the B.S.\ degree in electrical engineering from Tulane
  University and M.S.\ and Ph.D.\ degrees in electrical engineering
  from the University of Washington.  He is presently an associate
  professor in the School of Electrical Engineering and Computer
  Science at Washington State University.  His research interests
  include the use of computational methods to analyze acoustic,
  elastic, and electromagnetic wave propagation.
\end{biographynophoto}

\begin{biographynophoto}{Peter J.\ Bevelacqua}
  received the B.S. degree in Electrical Engineering (summa cum laude)
  from Washington State University in 2002.  In 2002 he worked as an
  intern at Sandia National Labs and was awarded a National Science
  Foundation fellowship in 2003.  He currently is a graduate student
  at Stanford University.
\end{biographynophoto}

\end{document}